\documentclass[draft]
  {aipproc}
\usepackage{amsmath}
\layoutstyle{6x9}

\begin{document}

\title{A practical model of convective dynamics for stellar evolution calculations}


\author{Neil Miller}{
  address={Department of Astronomy and Astrophysics, University of California, Santa Cruz}
}

\author{Pascale Garaud}{
  address={Department of Applied Mathematics and Statistics, University of California, Santa Cruz}
}

\begin{abstract}
Turbulent motions in the interior of a star play an important role in its evolution,
since they transport chemical species, thermal energy and angular momentum.
Our overall goal is to construct a practical turbulent closure model for convective
transport that can be used in a multi-dimensional stellar evolution calculation
including the effects of rotation, shear and magnetic fields.
Here, we focus on the first
step of this task: capturing the well-known transition from radiative
heat transport to turbulent convection with and without rotation, as well as the 
asymptotic relationship between turbulent and radiative transport in the limit of large
Rayleigh number.  We extend the closure model developed by \cite{Ogil03} and 
\cite{GarOgil05} to include heat transport and compare it with experimental results 
of Rayleigh-Benard convection.
\end{abstract}

\maketitle

\section{Introduction}
Turbulent convection plays an essential role in the evolution of most stars.
Turbulent motions are often the dominant mechanism of energy transport 
in stellar convective zones, and can increase momentum transport 
and chemical mixing by several orders of magnitude.
In many rotating astronomical systems where turbulence is anisotropic,
Reynolds stresses are the dominant transporters of angular momentum
and therefore influence the internal dynamics of the whole system.

It is currently not possible to perform a 3-D numerical simulation of
convective motions over the evolution timescale of a star.
However, when considering stellar evolution we are not necessarily interested in
the specific details of the convective motions, but rather in statistical 
properties such as the convective flux and Reynolds stresses.  
Our long-term goal is to construct a closed set of evolution equations for these 
statistical quantities in terms of 
large-scale system properties (e.g. viscosity, rotation, shear), which
can be used in a stellar evolution calculation.

In this paper we focus on modelling how the convective
Reynolds stresses and heat fluxes are affected by rotation. 
We are especially interested in approximately predicting the onset of
convection as well as the asymptotic behaviour of turbulent transport
for large Rayleigh number; capturing  
the onset is required for the stellar evolution 
model to correctly place the boundary between the convective 
and radiative zones.  Adequate prediction of the asymptotic behaviour of the convective 
turbulence is important to describe the amount of  energy and angular
momentum that is transported in the majority of the convective zone.

To quantify the quality of the proposed closure model 
we test it against linear stability analysis, numerical simulations and 
laboratory experiments of the rotating Rayleigh-Benard problem.  

\section{Rotating Rayleigh Benard Convection}
The typical Rayleigh-Benard convection setup is as follows:  
two rigid, ideally infinite, horizontal plates separated
by a distance $D$ confine a weakly compressible fluid between them.  
A temperature difference $\Delta T$ is maintained between a hot bottom and a cool top.  
The system is assumed to be rotating
with average angular velocity $\bar{\Omega} = \Omega \hat{z}$ with gravity ${\bf g} = -g\hat{z}$.  
The Boussinesq approximation is valid in this system.  
The governing equations are
\begin{eqnarray}
  \partial_{i} u_i &=& 0, \\
  (\partial_t + u_k \partial_k) u_i + 2 \epsilon_{ijk} \Omega_j u_k
     &=& -\alpha \Theta g_i - \partial_i \Psi + \nu \partial_{kk} u_i, \\
  (\partial_t + u_k \partial_k) \Theta &=& \kappa \partial_{kk} \Theta
\end{eqnarray}
where the dynamical variables are the temperature offset $\Theta$, 
the pressure perturbation from hydrostatic equilibrium $\Psi$
and the flow velocity ${\bf u}$.  The following parameters
are assumed to be constant: the coefficient of expansion $\alpha$,
the kinetic viscosity $\nu$ and the thermal diffusivity $\kappa$.
Sums over repeated indices are implied.

The qualitative behaviour of the system is controlled by three dimensionless quantities: 
the Rayleigh number, $\textrm{Ra} \equiv \alpha g \Delta T D^3 / (\kappa \nu)$ 
(measuring the ratio between buoyancy forces to viscous stabilising
forces), the Taylor number, $\textrm{Ta} \equiv 4 D^4 \Omega^2 / \nu^2$
(measuring the ratio between centrifugal forces to viscous forces), 
and the Prandtl number, $\textrm{Pr} \equiv \nu / \kappa$
(measuring the ratio between the viscous
diffusion rate and the thermal diffusion rate).  
For any given Taylor number and Prandtl number, there exists a critical 
Rayleigh number (Ra$_\textrm{c}$) above which the system
is convective and below which the system is conductive.
We strive to construct our model to match the known variation of the critical 
Rayleigh number with Taylor number and Prandtl number.  We also verify
that at Rayleigh numbers much larger than critical
the Nusselt number Nu (ratio of total heat flux to conductive heat flux) 
is correctly predicted by the model.

\section{Closure Model}
Our goal is to capture the behaviour of turbulent convection in a rotating system
both in terms of the onset of turbulence and of its asymptotic properties.
We developed a second order closure model  
using the technique described by \cite{Ogil03} and \cite{GarOgil05}.

We write each quantity as the sum of a mean and fluctuating part
($u = \bar{u} + u'$, $\Psi = \bar{\Psi} + \Psi'$,
and $\Theta = \bar{\Theta} + \Theta'$) -- in the Rayleigh-Benard
problem, these mean quantities only vary with $z$.

We define the following correlation quantities: $R_{ij} = \overline{ u_i' u_j' }$,
$F_i = \overline{ \Theta' u_i' }$, and $Q = \overline{ \Theta' \Theta' }$,
so that $R = R_{ii}$ is twice the mean turbulent kinetic energy.

The exact equations governing the mean quantities are
\begin{align}
\partial_i \bar{u}_i = 0 \label{cl1}\\
(\partial_t + \bar{u}_k \partial_k ) \bar{u}_i + 2 \epsilon_{ijk} \Omega_j \bar{u}_k 
   = - \alpha \bar{\Theta} g_i - \partial_i \bar{\Psi} 
     + \nu \partial_{kk} \bar{u}_i-\partial_j R_{ij} \label{cl2} \\
(\partial_t + \bar{u}_k \partial_k) \bar{\Theta} 
   = \kappa \partial_{ii} \bar{\Theta} - \partial_k F_k \label{cl3}
\end{align}
while those governing the second order correlation terms are modelled as
\begin{align}
  (\partial_t+\bar u_k\partial_k)R_{ij}+
  R_{ik}\partial_k\bar u_j+R_{jk}\partial_k\bar u_i 
  +2\epsilon_{ilm} \Omega_l R_{jm} + 2 \epsilon_{jlm} \Omega_l R_{im} 
  +\alpha(F_i g_j+F_j g_i) \nonumber\\ 
  - \nu \partial_{kk} R_{ij} =-C_1 \tau^{-1} R_{ij}-
  C_2 \tau^{-1} (R_{ij}-{\textstyle{\frac{1}{3}}}R\delta_{ij}) 
  - \nu C_{\nu} L^{-2} R_{ij},\label{cl4}\\
  (\partial_t+\bar u_j\partial_j)F_i+
  R_{ij}\partial_j\bar \Theta+ F_j\partial_j\bar u_i 
   +2\epsilon_{ijk} \Omega_j F_k+\alpha Qg_i - \frac{1}{2} (\nu + \kappa) \partial_{kk} F_i \nonumber\\
  =-C_6 \tau^{-1} F_i
   - \frac{1}{2} (\nu + \kappa) C_{\nu} F_i L^{-2}, \label{cl5} \textrm{     and}\\
  (\partial_t+\bar u_i\partial_i)Q+2F_i\partial_i\bar \Theta - \kappa \partial_{kk} Q
  = -C_7 \tau^{-1} Q - \kappa C_{\nu} Q L^{-2} \label{cl6}
\end{align}
where the left-hand-side of each equation is exact, while the right-hand-side
models the effect of higher order correlations.  
The constants $C_1, C_2, C_{\nu}, C_6, C_7$ are free parameters of the 
closure model\footnote{Note that $C_1\simeq 0.4,C_2\simeq0.6,$ and $C_{\nu}\simeq12$ have already been found to 
give an adequate description of the turbulent stresses in Couette-Taylor experiments by Garaud and Ogilvie (2005),
but $C_6$ and $C_7$ remain to be determined.}. 
The variable $\tau$ is the characteristic timescale for the redistribution of energy
along the turbulent cascade, which is controlled by the turnover time of the largest
eddies $d^{-1} R^{1/2}$ and 
$L$ is the characteristic size of the perturbations near onset, which we now describe in more detail.

Following Prandtl's mixing length theory, \cite{GarOgil05} originally 
suggested that in a wall-bounded experiment $L$ can be thought of as the distance
to the wall.  
However, rotation does no net work on the system when $\Omega$ is parallel to $\bf{g}$
and since the closure model is constructed on energetic arguments,
using this lengthscale prescription here fails to 
capture the known effects of rotation on the onset and turbulent properties of convection (\cite{Chan61}).

Physically, rotation influences the onset of convection by
decreasing the characteristic lengthscale of convective motions
in the direction perpendicular to the rotation axis. 
To capture this effect, we construct our lengthscale $L$ to be
the harmonic mean between the distance to the wall and the wavelength of
the most linearly unstable mode $\lambda$ which, in the Rayleigh-Benard experiment, is a function 
of the Taylor number: therefore 
\begin{equation}
  L = \left(\frac{1}{d^2}+\frac{2}{\lambda^2}\right)^{-1/2}.
  \label{lenscl}
\end{equation}

\section{Results}
We seek solutions to equations (\ref{cl1}) - (\ref{lenscl})
assuming no-slip boundary conditions and fixed uniform plate temperatures.
The no-slip boundary condition states that $u_i = 0$ at the boundaries, implying
$R_{ij} = F_i = 0$.  At the lower boundary $\Theta = \Delta T$
and at the upper boundary $\Theta = 0$.  Since the temperature perturbations are zero
at both boundaries, $Q = 0$.

\begin{figure}
  \includegraphics[]{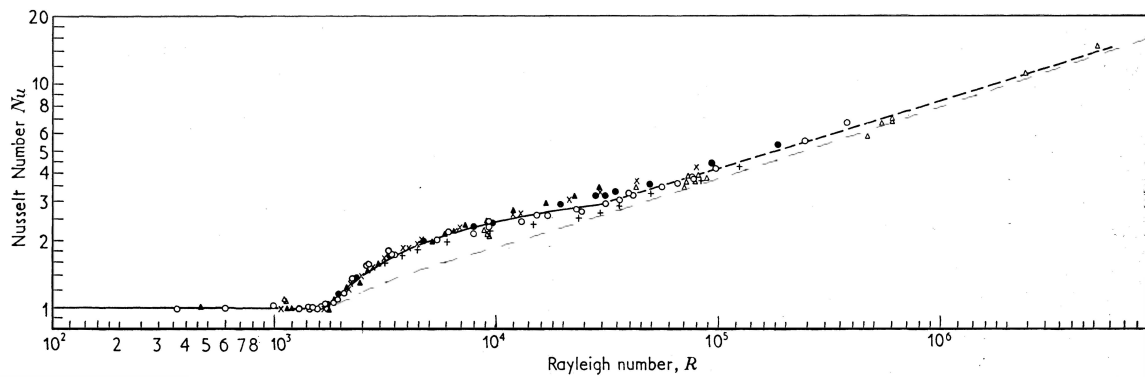}
  \caption{The Rayleigh-Nusselt relation adapted from \cite{Chan61}.  
  The dashed line is the model prediction for
  $\Omega=0, \nu=10^{-3}, \kappa=10^{-4}$ generated by varying the temperature difference
  between the two plates.}
\end{figure}

In Figure 1 we show the Nusselt number - Rayleigh number relationship 
in the non-rotating case for both our model and
the selection of experimental data summarised by \cite{Chan61}.
Note that our model shows good agreement with the experimental
data for the critical Rayleigh number where the transition between
conductive (Nu = 1) and convective (Nu > 1) heat transport occurs.
Our model also reproduces the standard power law relationship
$\textrm{Nu} \propto \textrm{Ra}^{1/3}$ at high Rayleigh number which is a natural consequence
of our selection $L \propto d$ when $\Omega=0$ (cf. Prandtl's mixing length theory).  
However, the good quantitative agreement between the experimental data and the model 
prediction for Nu was unexpected since
$C_6$ or $C_7$, which play a role when the system is convective,
have not yet been adjusted from their default value of unity.  By adjusting
these we should be able to further improve the correspondence of the model to reality
for $\textrm{Ra} > \textrm{ Ra}_{\textrm{c}}$.

\begin{figure}
  \includegraphics[]{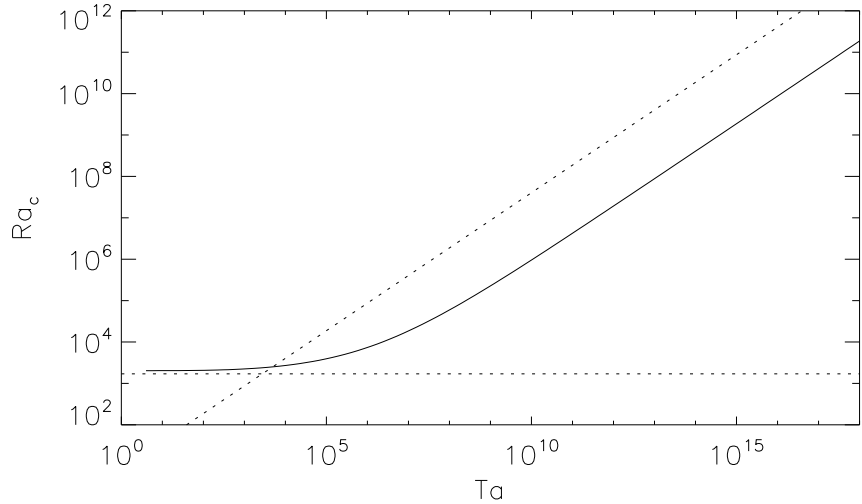}
  \caption{The Taylor-Rayleigh critical relation.  The solid line is the model prediction for the
  critical Rayleigh number  as a function of Taylor number with $\nu=10^{-3}, \kappa=10^{-4}$.
  The dotted lines are the asymptotic relations derived from linear stability analysis by \cite{Chan61}
  for $\Omega=0$ and for $\textrm{Ta}\rightarrow \infty$ for the direct mode of instability
  (see equation 184 on page 106).} % page 95
\end{figure}

In rotating systems, convection is known to delay of the onset of convection at high Taylor numbers.  
Linear theory (cf. \cite{Chan61}) predicts that $\textrm{Ra}_{\textrm{c}} \propto \textrm{Ta}^{2/3}$,
with a coefficient of proportionality which depends somewhat on the boundary conditions
and on the Prandtl number of the system.
In Figure 2, we compare this known relationship to our model (solid line).  
The predicted power law matches linear theory, but the coefficient of
proportionality is somewhat smaller than required.
Nonetheless, we consider the agreement satisfactory considering the simplicity
of this closure model.

To conclude, we find that the closure model adequately describes both the turbulent 
convective heat flux as a function of Rayleigh number in the absence of rotation
as well as the delay of the onset of convection in a rotating system.  
In future work, we intend to compare the degree to which the model matches
numerical simulations of developed convection in the rotating Rayleigh-Benard problem 
(e.g. \cite{Jul96}). 
This comparison will be useful in determining the utility of our model
away from the onset of convection in a rotating system and permit the calibration of
$C_6$ and $C_7$, the remaining free parameters of the model.

\begin{theacknowledgments}
  N. Miller and P. Garaud gratefully acknowledge funding from
  NSF-AST-0607495.  We thank Gordon Ogilvie for his guidance
  throughout the completion of this work.
\end{theacknowledgments}

\bibliographystyle{aipproc}   
\bibliography{myBib}

\IfFileExists{\jobname.bbl}{}
 {\typeout{}
  \typeout{******************************************}
  \typeout{** Please run "bibtex \jobname" to optain}
  \typeout{** the bibliography and then re-run LaTeX}
  \typeout{** twice to fix the references!}
  \typeout{******************************************}
  \typeout{}
 }

\end{document}